\documentclass[twocolumn,showpacs,aps,amssymb,floatfix,prd,amsmath,preprintnumbers]{revtex4}
\usepackage{epstopdf}
\usepackage{capt-of}
\usepackage{graphicx}  
\usepackage{dcolumn}   
\usepackage{bm}
\begin{document}
\input epsf.tex

\title{Einstein energy-momentum complex for a phantom black hole metric\footnote{ \scriptsize This collaborative research work was done during a  workshop on  Introduction to Research in Einstein's General Relativity at NIT, Patna (India). Authors' permanent addresses are given below.}}

\author{
    P. K. Sahoo\footnote{\tiny Department of Mathematics, Birla Institute of Technology and Science-Pilani, Hyderabad Campus, Hyderabad 500078, Andhra
                        Pradesh, India,  Email:  sahoomaku@rediffmail.com.},
    K. L. Mahanta\footnote{\tiny Department of Mathematics, C.V. Raman College of Engineering, Bhubaneswar 752054, Odhisa, India, Email: kamal2\_m@yahoo.com.},
    D. Goit\footnote{ \tiny Department of Physics, B.S. College, Danapur, Patna-800012, Bihar, India, E-mail:  goitdn@gmail.com.},
    A. K. Sinha\footnote{\tiny Department of Physics, College of Commerce, Kankarbagh, Patna 800020, Bihar, India, e-mail: ashutosh25june@gmail.com.},
    S. S. Xulu \footnote{\tiny Department of Computer Science, University of Zululand,3886 Kwa-Dlangezwa,South Africa. e-mail:ssxulu@pan.uzulu.ac.za.},
    U. R. Das\footnote{\tiny Department of Physics, College of Commerce, Kankarbagh, Patna 800020, Bihar, India, e-mail: Upendraphy342@gmail.com.},
    A. Prasad\footnote{\tiny Department of Physics, D.N. College, Masaurhi, Patna 804452, Bihar, India, e-mail: arbindprasad57@gmail.com.} and
    R.  Prasad\footnote{\tiny Department of Physics, L.S. College, Muzaffarpur 842001, Bihar, India, e-mail: pd.rajendra.phy@gmail.com.}}

\affiliation{National Institute of Technology, Patna 800005, Bihar, India.}

\begin{abstract}
In this paper we  calculate  the energy distribution $E(r)$ associated with a static spherically symmetric non-singular phantom black hole metric in Einstein's prescription in  general relativity. As required for the Einstein energy-momentum complex, we perform calculations in quasi-Cartesian coordinates. We also calculate momentum components and get zero values as expected from the geometry of the metric.

\end{abstract}




\maketitle


\section{Introduction}

Since the beginning of the Einstein's theory of general relativity (GR), there are several known important issues; e.g.,
the missing matter cosmology problem and the energy-momentum localization in curved spacetimes,
that are still in doubt and possess non-specific solutions. Nevertheless, there is a wide range of
attempts for several researchers to beat these problems, using various hypotheses and  tools. In GR (that is, in curved spacetimes),
the partial derivative of the usual local conservation equation $T^k_{i,k}=0$ valid in Minkowski spacetime is replaced by a covariant derivative.
$T^k_i$, the energy-momentum  tensor of matter and all non-gravitational fields, no longer satisfies $T^k_{i,k}=0$ in presence of gravitational field.
The contribution from  the gravitational field is now required to construct an energy-momentum expression which satisfies a local conservation law.

Einstein solved this problem and suggested an expression for energy-momentum distribution (for a great detail, see in \cite{Moller}).
Despite his work was criticized  by a few physicist (e.g., notably by Pauli), he justified that his energy-momentum complex provides convincing
results for the total energy and momentum of isolated systems. Later, many physicists including Tolman\cite{Tolman},
Landau and Lifshitz\cite{LL}, Weinberg\cite{Weinberg},  Papapetrou\cite{Papapetrou}, and Bergmann and Thomson\cite{BT} suggested alternative
expressions for the energy-momentum distribution. The main problem with all these definitions is that they are coordinate-dependent.
One can  however obtain meaningful results for the total energy and momentum of isolated systems only when calculations are performed in
quasi-Cartesian coordinates. These complexes are also called pseudotensors because they are not tensorial objects.
Komar\cite{Komar} and Penrose\cite{Penrose} though constructed coordinate-independent  definitions of energy and angular momentum,
their approach were restricted  only to a limited class of metrics. Therefore, the coordinate-independent approaches did in fact worse.

In 1990, Virabhadra's  seminal papers  shook the notion that  energy-momentum complexes could give sensible results {\em only for}
the total energy of isolated systems.  Virbhadra\cite{Virbhadra1990} showed that several energy-momentum complexes give the same
energy-distribution for the  Kerr-Newman metric. In this context, Virbhadra and his collaborators\cite{VirbhadraColl}
studied many spacetimes and obtained energy distributions for those.
Nathan Rosen (an eminent  collaborator of Albert Einstein) and Virbhadra\cite{RoseVir} studied energy and momentum
distributions in Einstein-Rosen  cylindrically  gravitational waves. To a great surprise, several complexes produced
same results though this metric is not asymptotically flat.
Aguirregabiria {\it et al.}\cite{ACV} showed that several coordinate-dependent definitions give the same energy and  momentum
 distribution for any metric of Kerr-Schild class. In 1999, Virbhadra\cite{Virbhadra1999} proved that various energy-momentum complexes
   coincide not only for the Kerr-Schild class metrics, but for a class of solutions much more general than that. This includes
   asymptotically flat as well as non-flat spacetimes.

The problem of energy-momentum localization in GR gained a new point of view with the result studied by Virbhadra and his collaborators.
Using Einstein  energy-momentum complex, Rosen\cite{Rosen} found that the total energy is zero for a closed  homogeneous isotropic universe
described by a Friedmann-Robertson-Walker (FRW) metric. Johri {\it et al}. \cite{Johri}, using the Landau and Lifshitz energy-momentum complex,
showed that the  total energy of an FRW spatially closed universe is zero at all times irrespective of the equations of  state of the cosmic fluid. They also showed that the total energy enclosed within any finite volume of the  spatially flat FRW universe is zero at all times.

Many authors tried to solve energy-momentum localization problem using different spacetimes and different energy-momentum prescriptions and
they obtained a plethora of important results (see \cite{Many1,Many2,Many3} and references therein).  Xulu  studied several spacetimes (asymptotically flat as well as non-flat).  He obtained the  energy distribution of a charged dilaton black hole and  Melvin's magnetic universe. Papapetrou and Weinberg complexes for the anisotropic Bianchi type I space time were investigated by Xulu.

Radinschi used Landau-Lifshitz and Papapetrou energy-momentum solutions for Bianchi type $VI_0$ spacetime. Later, she obtained results
for the same  metric using Tolman, Bergmann-Thomson and M\o ller energy-momentum complexes. Loi and Vargas  studied energy localization
for Bianchi I and II universes in teleparallel gravity. Aydogdu investigated Einstein and Landau and Lifshitz energy-momentum complexes for
Bianchi type-II universe in GR. Aydogdu and Salti  used  Einstein  Bergmann-Thomson prescriptions for Bianchi type-V metric in general
relativity and teleparallel gravity. Andrade {\it et al.}\cite {Andrade2000} obtained a conserved energy-momentum gauge current in the
context of a gauge theory for translation group.

Recent analysis of type Ia supernovae, cosmic microwave background anisotropy and mass power spectrum observations favor the negative
values of the equation of state parameter $\omega$  for the dark energy \cite{Many4}. Considering the equation of state parameter of
accustomed quintessence models with positive kinetic energy it is not possible to derive the aforesaid order of $\omega$. Therefore,
many authors \cite{Many5} studied the phantom field models with negative kinetic energy to achieve this regime of $\omega$.
If this candidate of dark energy is part of the real field content of the large scale structure of the Universe,
it is natural to look for its manifestation in the study of black holes. The exact solution of
black holes in phantom field is called phantom black holes. Gao and Zhang \cite{Gao} discussed the cosmological aspects of the phantom
black hole and phantom field. Babichev {\it et al}. \cite{Babi} studied the accretion of phantom fluid onto a black hole.
Bronnikov and Fabris investigated the physics of neutral phantom black holes and presented some interesting results \cite{Bron}. Ding {\it et al}.\cite{Ding} studied the influence of phantom fields on strong gravitational lensing.

The purpose of this paper is to calculate the energy and momentum distributions in a phantom  black hole spacetime in  Einstein's prescription.
Here, we use the convention that Latin indices take values from 0 to 3 and Greek indices run from 1 to 3. As usual in general relativity papers,
we also use  $G=1, c = 1$ units.
\section{Phantom black hole metric}

The Bronnikov-Fabris phantom black hole metric\cite{BF}, expressed by Ding {\it et al}.\cite{Ding} in a neat form,  is

\begin{equation}
ds^2= f(r) dt^2 - \frac{dr^2}{f(r)} - (r^2+p^2) (d\theta^{2}+\sin^{2}\theta d\phi^2)
\end{equation}
with
\begin{equation}
f(r)=1-\frac{3M}{p}\biggl[\biggl(\frac{\pi}{2}-\arctan \frac{r}{p}\biggr)\biggl(1+\frac{r^2}{p^2}\biggr)-\frac{r}{p}\biggr] \text{,}
\label{f}
\end{equation}
where $M$ is a mass parameter defined in the usual way and  $p$ is a positive constant termed as the phantom constant\cite{Ding} (Symbol $p$ is meant for phantom.)
Ding {\it et al.}\cite{Ding} explained that for $M=0$, the metric represents Ellis wormhole.

In order to calculate energy and momentum components, we  transform the line element (1)  to quasi-Cartesian coordinates $t, x, y, z$ using
the following transformation:
\begin{eqnarray}
x &=& r \sin \theta \cos \phi \text{,} \nonumber \\
y &=& r \sin \theta \sin \phi  \text{,} \nonumber \\
z &=& r \cos \theta \text{.}
\end{eqnarray}
The line element (1) becomes
\begin{eqnarray}
ds^{2}&=&f(r) dt^{2}- \frac{r^2+p^2}{r^2}(dx^2+dy^2+dz^2) \nonumber  \\
      &&-\biggl(\frac{1}{f(r)}-\frac{r^2+p^2}{r^2}\biggr) \biggl(\frac{xdx+ydy+zdz}{r}\biggr)^2
\end{eqnarray}
where
\begin{equation}
r = \sqrt{x^2+y^2+z^2} \text{.}
\end{equation}
The determinant of the metric tensor $g_{ik}$ is
\begin{equation}
g = - \left(1+\frac{p^2}{r^2}\right)^2 \text{.}
\end{equation}
and 10 independent contravariant components of  the symmetric metric tensor ($g^{ik} = g^{ki}$  for all values of $i,k$) are
\begin{eqnarray}
g^{00} &=& \frac{1}{f}  \text{,}\nonumber \\
g^{11} &=&\frac{- \left(p^2+r^2\right) f x^2-\left(r^2-x^2\right) r^2}{\left(p^2+r^2\right) r^2} \text{,}\nonumber\\
g^{22} &=&\frac{-\left(p^2+r^2\right) fy^2 -\left(r^2-y^2\right) r^2}{\left(p^2+r^2\right) r^2}\text{,}\nonumber \\
g^{33} &=& \frac{-\left(p^2+r^2\right) fz^2 -\left(r^2-z^2\right) r^2}{\left(p^2+r^2\right) r^2}\text{,}\nonumber \\
g^{01} &=& 0 \text{,}\nonumber \\
g^{02} &=& 0 \text{,}\nonumber \\
g^{03} &=& 0 \text{,}\nonumber \\
g^{12} &=& \frac{\left(r^2-\left(p^2+r^2\right) f\right)xy}{\left(p^2+r^2\right) r^2}  \text{,}\nonumber \\
g^{23} &=& \frac{\left(r^2-\left(p^2+r^2\right) f\right)yz}{\left(p^2+r^2\right) r^2}  \text{,}\nonumber \\
g^{31} &=& \frac{\left(r^2-\left(p^2+r^2\right) f\right)xz}{\left(p^2+r^2\right) r^2}   \text{.}
\end{eqnarray}
\section{Einstein energy-momentum definition in GR}

A thorough study by Virbhadra \cite{Virbhadra1999} revealed that the Einstein energy-momentum complex gives the most reliable
energy distribution and therefore we will use the same definition in this paper. The energy-momentum complex of Einstein is \cite{Moller}
\begin{equation}
\Theta^k_i=\frac{1}{16\pi}H^{kl}_{i,l},
\end{equation}
where
\begin{equation}
H^{kl}_i=-H^{lk}_i=\frac{g_{in}}{\sqrt{-g}}\biggl[-g\biggl(g^{kn}g^{lm}-g^{ln}g^{km}\biggr)\biggr]_{,m}.
\end{equation}
$\Theta^0_0$ and $\Theta^0_{\alpha}$ denote for the energy and momentum density components, respectively.
(Virbhadra\cite{Virbhadra1999} mentioned that though the energy-momentum complex found by Tolman differs in
form from the Einstein energy-momentum complex, both are equivalent in import.)

$\Theta^k_i$ satisfies the covariant local conservation laws:
\begin{equation}
\frac{\partial \Theta_i{}^{k}}{\partial x^k} = 0 \text{.}
\end{equation}
The energy-momentum components are expressed as
\begin{equation}
P_i=\int \int \int \Theta^0_i \ dx^1dx^2dx^3.
\end{equation}
$P_{\alpha}$ gives momentum components $P_1$, $P_2$, $P_3$ and $P_0$ gives the energy. Using Gauss's theorem in above, one can get

\begin{equation}
P_i=\frac{1}{16\pi} \int \int H^{0\alpha}_{i} \ \eta_{\alpha} \ dS,
\label{Pi}
\end{equation}
where $\eta_{\alpha}$ is the outward unit normal vector over the infinitesimal surface element $dS$.
In order to obtain energy, we  obtain only 3 components of $H^{kl}_i$:
\begin{eqnarray}
H^{01}_0 &=&  \frac{2x\left(fp^2-fr^2+r^2\right)} {r^4}  \text{,}\nonumber \\
H^{02}_0 &=&  \frac{2y\left(fp^2-fr^2+r^2\right)} {r^4}  \text{,}\nonumber \\
H^{03}_0 &=&  \frac{2z\left(fp^2-fr^2+r^2\right)} {r^4}      \text{.}
\label{H}
\end{eqnarray}
We use $(\ref{H})$ in $(\ref{Pi})$ and get the energy distribution:
\begin{equation}
E(r)= \frac{f \left(p^2-r^2\right)+r^2}{2r} \text{,}
\label{Energy}
\end{equation}
where $f$ is defined in equation $(\ref{f})$. $E(r)$ is  the total (matter plus  gravitational field) energy  within radius $r$. Similarly, momentum is the   total momentum due to matter and gravitational field both.

Similarly, in order to obtain momentum components, we calculate $H^{01}_1, H^{02}_1,  H^{03}_1$, $H^{01}_2,  H^{02}_2, H^{03}_2$,
$H^{01}_3,  H^{02}_3$,  and  $H^{03}_3$.
\begin{eqnarray}
H^{01}_1&=& H^{02}_1=H^{03}_1= 0 \text{,} \nonumber \\
H^{01}_2&=&H^{02}_2= H^{03}_2= 0 \text{,}  \nonumber \\
H^{01}_3&=&H^{02}_3=H^{03}_3=0    \text{.}
\label{HM}
\end{eqnarray}

We use  equation $(\ref{HM})$ in equation $(\ref{Pi})$. As expected for a static metric, we get momentum components:
\begin{eqnarray}
P_x  &=& 0     \text{,}\nonumber \\
P_y  &=& 0     \text{,}\nonumber \\
P_z  &=& 0     \text{.}
\label{Momentum}
\end{eqnarray}

\begin{figure}[!h]
\centering
\includegraphics[scale=0.65]{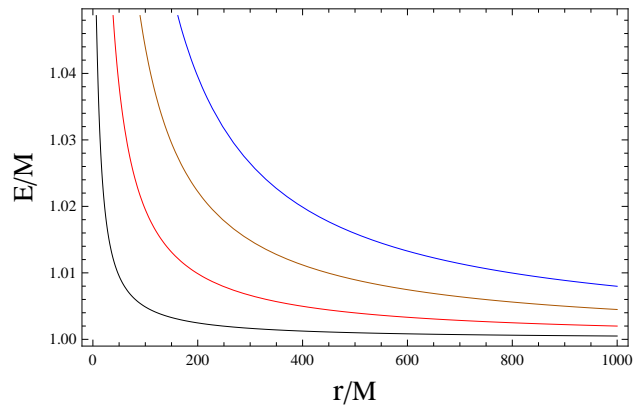}
\caption{This figure shows the ratio of the energy to the mass  $E/M$ vs. the ratio of the radial distance to the mass $r/M$
for several values of  the ratio of phantom constant to the mass parameter $p/M = 1$ (black), $2$ (red), $3$(orange),
and $4$ (blue). As $r/M$ approaches infinity, $E/M$ tends to  $1$.}
\label{fig1}
\end{figure}

\begin{figure}[!h]
\centering
\includegraphics[scale=0.4]{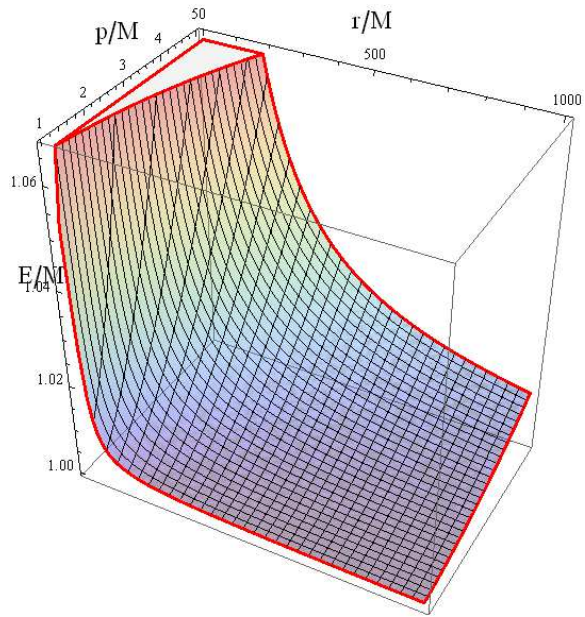}
\caption{In this 2-dimensional surface plot, the ratio of the energy to the mass parameter $E/M$ is plotted against the
ratio of the radial distance to the mass $r/M$  and the ratio of phantom constant to  the mass parameter $p/M$.}
\label{fig2}
\end{figure}


Now, we plot graphs (refer to  figures 1 and 2) to analyze the nature of energy distribution $E(r)$
[see equation $(\ref{Energy})$] as the radial distance and phantom constant increase while we keep  the mass  parameter fixed. In figure 1,
we plot the  ratio of the energy to the mass  $E(r)/M$  against the ratio of the radial distance to the mass $r/M$  for 4  different values of
the ratio of phantom  constant and the mass parameter $p/M$. We maintain the total mass parameter $M$ constant.
These curves have the same horizontal asymptote $E/M = 1$. This tells that as $r \rightarrow \infty$, $E(r) \rightarrow M$.
If we maintain $p$ constant, the energy content $E(r)$ decreases  with decrease in the radial distance $r$.
This exhibits that phantom field has negative energy. We further  find that, with fixed $r/M$, $E/M$ is bigger for larger  value of phantom constant.
It is very exciting to note that the decrease rate of $E(r)$ with increase in $r$ is higher for smaller values of the phantom constant.
In  surface plot (refer to figure 2), these results are exhibited more elegantly.

\section{Discussion and Conclusion}
There have been some different considerations concerning with spherically symmetric systems in the framework of energy momentum localization.
Misner {\em et al} (see in \cite{Mis}) concluded that energy can be localized only for spherically symmetric  systems.
Cooperstock and Sarracino\cite{Coop} however opposed that idea and  stated  that  localizability of energy  cannot  depend on gemetry of spacetime.
The energy-momentum complexes are non-tensorial under general coordinate transformations and  are restricted to   their uses in quasi-Cartesian  coordinates only.
Pioneered by Virbhadra, numerous scientists from all over the world did a lot of work showing that energy-momentum complexes are indeed
very useful tools in general relativity.

One could ask why should we study energy distribution in a spacetime. The answer is that, by knowing energy-momentum distributions,
we get an excellent idea of the spacetime. As it is already discussed by others, it gives a good idea of effective gravitational mass of the
  object causing spacetime curvature. In addition, it gives an intuitive feeling about the gravitational lensing in that spacetime.
  Negative energy region  is likely to serve as a divergent lens and positive energy region  as a convergent lens.
  The analysis of energy distributions in spacetimes helped Virbhadra  discover excellent lensing phenomena\cite{VirEllisLens}.
  For illustration, our  analysis in this paper proves that phantom causes negative energy and this is why when we increase $r$,
  the  energy content  $E(r)$ decreases.  Thus, our analyses indicate that phantom field  would cause  repulsive effects to causal geodesics.
  This is really an exciting and important result.

In this paper, we calculated  energy and momentum distributions in the Einstein prescription for phantom black hole metric in quasi-Cartesian coordinates.
Further work towards the  investigation of the energy-momentum for the phantom black hole spacetime is required.  Other prescriptions must be used and compared.
These calculations are very lengthy and time taking, and work is in progress.
\acknowledgments
 The authors thank the National Institute of Technology, Patna, India for organizing  and giving us opportunity  to
 attend the  {\em Workshop on Introduction to Research in Einstein's General Relativity} during which this research work was done.
 Also, the authors would like to thank   A. De (Director, NIT, Patna), N. Lall (Head, physics department, NIT, Patna),
 and the workshop organizer B. K. Sharma for their hospitality. PKS thanks the Institute of Mathematical Sciences   (IMSc), Chennai (India) for
 providing with facility and support during a visit where part of this work was done. SSX thanks the University of Zululand (South Africa)
 for all support where he carried out this work.

We are also  very indebted to the  editor and the anonymous referee of this journal for illuminating suggestions that have significantly
improved our paper in terms of  research  quality as well as the presentation.


\end{document}